\documentclass{PoS}

\title{The future Gamma-Ray Burst Mission {\it SVOM}}

\ShortTitle{{\it SVOM}, a future Mission for Gamma-Ray Burst Studies}

\author{
\speaker{S. Schanne}$^{1}$, J. Paul$^{1,8}$, J. Wei$^{2}$, S.-N. Zhang$^{3}$, S. Basa$^{4}$, J.-L. Atteia$^{5}$, D. Barret$^{6}$, A. Claret$^{1}$, B. Cordier$^{1}$, F. Daigne$^{7}$, P. Evans$^{9}$, G. Fraser$^{10}$, O. Godet$^{6}$,  D. Götz$^{1}$, P. Mandrou$^{6}$, J. Osborne$^{9}$ \\
$^{1}$ CEA Saclay, DSM/IRFU/Service d'Astrophysique, 91191, Gif-sur-Yvette, France\\
$^{2}$ National Astronomical Observatories/CAS, 20A Datun Road, Beijing 100012, China \\
$^{3}$ Institute of High Energy Physics/CAS, 19B YuquanLu, Beijing, 100049, China \\
$^{4}$ Laboratoire d'Astrophysique de Marseille, 38 rue Joliot-Curie, 13388 Marseille, France \\
$^{5}$ Laboratoire d'Astrophysique de Toulouse-Tarbes, 14 av. Belin, 31400 Toulouse, France \\
$^{6}$ Centre d'Etude Spatiale des Rayonnements, 9 av. du Colonel Roche, 31028 Toulouse, France \\
$^{7}$ Institut d'Astrophysique de Paris, 98bis bd. Arago, 75014 Paris, France \\
$^{8}$ Laboratoire Astroparticules et Cosmologie, 10 rue Domon et Duquet, 75205 Paris, France \\
$^{9}$ Dept. of Physics \& Astronomy, University of Leicester, Leicester LE1 7RH, UK \\
$^{10}$ Space Res. Centre, Dept. of Physics \& Astronomy, Univ. of Leicester, Leicester LE1 7RH, UK
}

\abstract{
The Space-based multi-band astronomical Variable Object Monitor (SVOM) is a future satellite mission for Gamma-Ray Burst (GRB) studies, developed in cooperation between the Chinese National Space Agency (CNSA), the Chinese Academy of Science (CAS), the French Space Agency (CNES) and French research institutes.
	The scientific objectives of the {\it SVOM} GRB studies cover their classification (GRB diversity and unity of the model), their physics (particle acceleration and radiation mechanisms), their progenitors, cosmological studies (host galaxies, star formation history, re-ionization, cosmological parameters), and fundamental physics (origin of cosmic rays, Lorentz invariance, gravitational wave sources).
	From 2015 on, {\it SVOM} will provide fast and accurate localizations of all known types of GRB, and determine the temporal and spectral properties of the GRB emission, thanks to a set of four onboard instruments.
	The trigger system of the coded-mask telescope ECLAIRs onboard SVOM images the sky in the 4-120 keV energy range, in order to detect and localize GRB in its 2 sr-wide field of view. The low-energy threshold of ECLAIRs is well suited for the detection of highly red-shifted GRB. 
	The high-energy coverage is extended up to 5 MeV thanks to a non-imaging gamma-ray spectrometer. 		GRB alerts are sent in real-time to the ground observers community, and a spacecraft slew is performed in order to place the GRB within the narrow fields of view of a soft X-ray telescope and a visible-band telescope, to refine the GRB position and study its early afterglow. 
	Ground-based robotic telescopes and wide-angle cameras complement the onboard instruments. 
	A large fraction of GRB will have redshift determinations, thanks to an observing strategy optimized to facilitate follow-up observations by large ground-based spectroscopic telescopes. 
	This paper presents an overview of the SVOM mission, its instruments and status.

}

\FullConference{The Extreme sky: Sampling the Universe above 10 keV - extremesky2009,\\
		October 13-17, 2009\\
		Otranto (Lecce) Italy}

\begin{document}

\section{Introduction}
% ------------------------------
	Forty years after its discovery, the Gamma-Ray Burst (GRB) phenomenon is not yet completely understood \cite{piran05}. 
	The cosmological nature of these transient sources of gamma-rays has been established, and their association with explosions of massive stars (>30 M$_{\odot}$) is a scenario that reproduces most observations, at least for long duration GRB.
	They have been detected up to redshift 8.2 \cite{tanvir09}, with the possibility to investigate the early Universe (star formation history, re-ionization era), and to derive cosmological parameters.
	Conversely for short duration GRB, mostly described by coalescences of two compact objects (black holes, neutron stars or white dwarfs), the situation is less consensual and lacks a good sample of  afterglow observations.
	Open questions concern the physical processes during the prompt phase (particle acceleration, radiation), the GRB classification, the characterization of GRB host galaxies and progenitors, as well as fundamental physics issues like Lorentz invariance, the origin of cosmic rays and gravitational waves.

\begin{figure}[h]
  \includegraphics[page=1,height=.20\textheight]{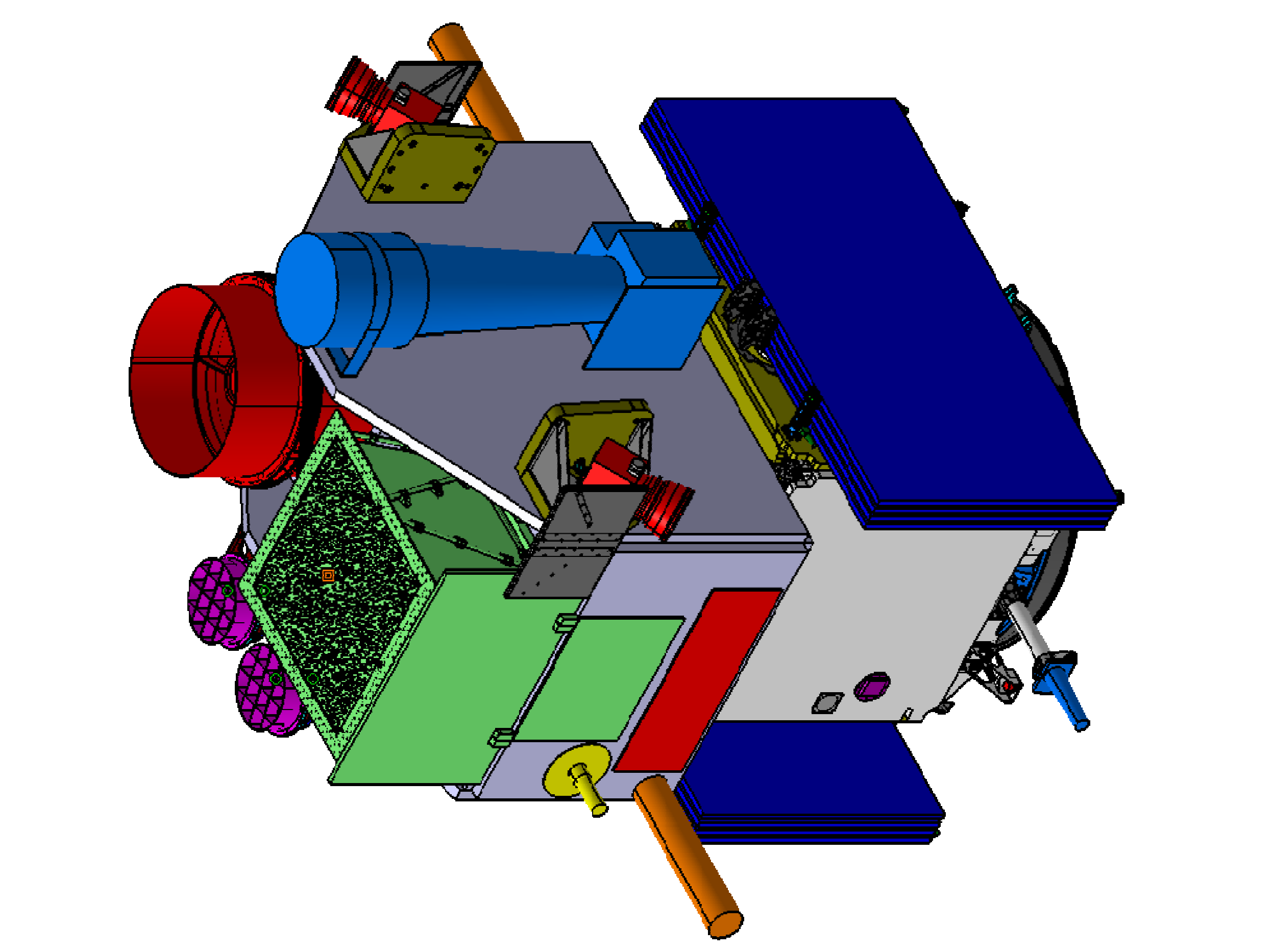}
\hspace{2cm}
   \includegraphics[page=2,height=.20\textheight]{paper-prepare-Ottranto.pdf}
 \caption{Left: sketch of a possible SVOM implementation, with the four instruments : ECLAIRs (green), MXT (blue), VT (red) and two GRM modules (purple). Right: SVOM multi-wavelength coverage.}
\label{fig:svom}
\end{figure}

	In order to contribute to the field of GRB research, the French Space Agency (CNES), the Chinese Space Agency (CNSA) and research laboratories from both countries are developing the {\it SVOM} mission (Space-based multi-band astronomical Variable Object Monitor).
	{\it SVOM} has successfully concluded its feasibility study (phase A) in 2008, and is expected to start its detailed design (phase B) in 2010, for a launch in 2015, on a circular orbit with an inclination of $\sim$30$^{\circ}$ and altitude of $\sim$600 km.
	Fig. \ref{fig:svom} shows a sketch of the {\it SVOM} satellite, which carries four instruments: (i) ECLAIRs (a coded-mask wide-field telescope for GRB real-time localizations to few arcmin), (ii) two GRM modules (non-imaging $\gamma$-ray spectrometers), and two narrow-field instruments, (iii) MXT and (iv) VT, with few arcsec-level localizations, for the study of the GRB early afterglow in X-ray and visible bands. 
	For this purpose, a GRB detection within the ECLAIRs wide field-of-view initiates an autonomous satellite slew to place the source in the narrow fields-of-view of the MXT and VT.
	The {\it SVOM} pointing strategy \cite{cordier08} avoids bright galactic X-ray sources; it is optimized for GRB detections towards the night sky on Earth, to enhance follow-up possibilities by large ground-based telescopes, with a goal of 75\% of GRB observable from ground during the early afterglow phase, at an expense of a mean $\sim$27\%-occultation of the ECLAIRs field-of-view by the Earth.

The space-based instruments of the {\it SVOM} mission are complemented by ground-based telescopes: (i) GWAC, wide-field visible-band cameras covering a large portion of the ECLAIRs field-of-view (based in China and following the ECLAIRs pointings) to catch the GRB prompt visible-band emission, and (ii)
GFT, two robotic telescopes (one based in China and one provided by France) which measure the photometric properties of the early afterglow from the near-infrared to the visible band, and refine the on-board GRB position. 
	Fig. \ref{fig:svom} summarizes the multi-wavelength capabilities of SVOM. 
	The following sections give details on the mission and its instruments.

\section{Space-based instruments}
% ------------------------------

\begin{figure}[ht]
\includegraphics[page=3,height=.22\textheight]{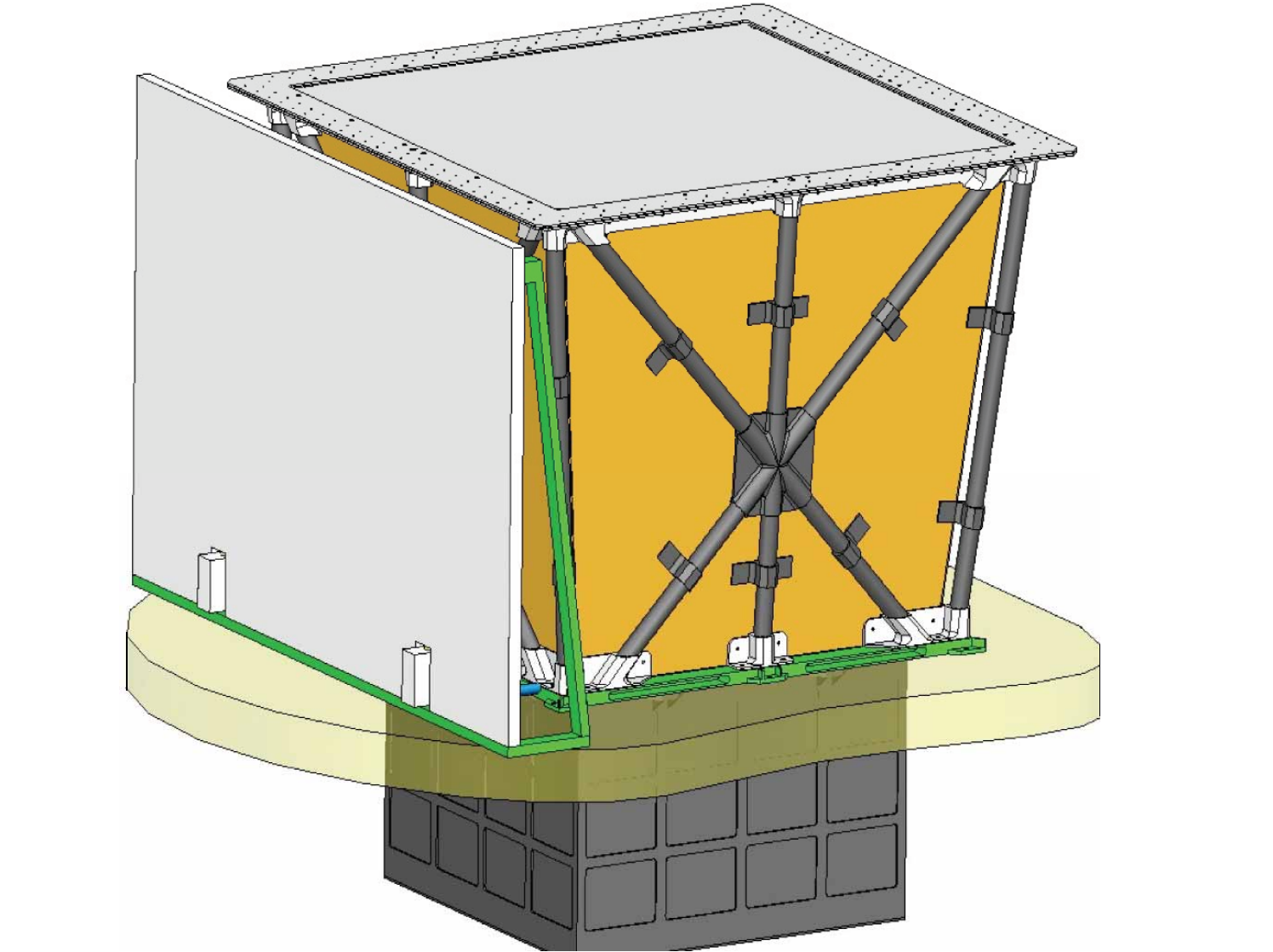}
\hspace{2cm}
  \includegraphics[page=4,height=.22\textheight]{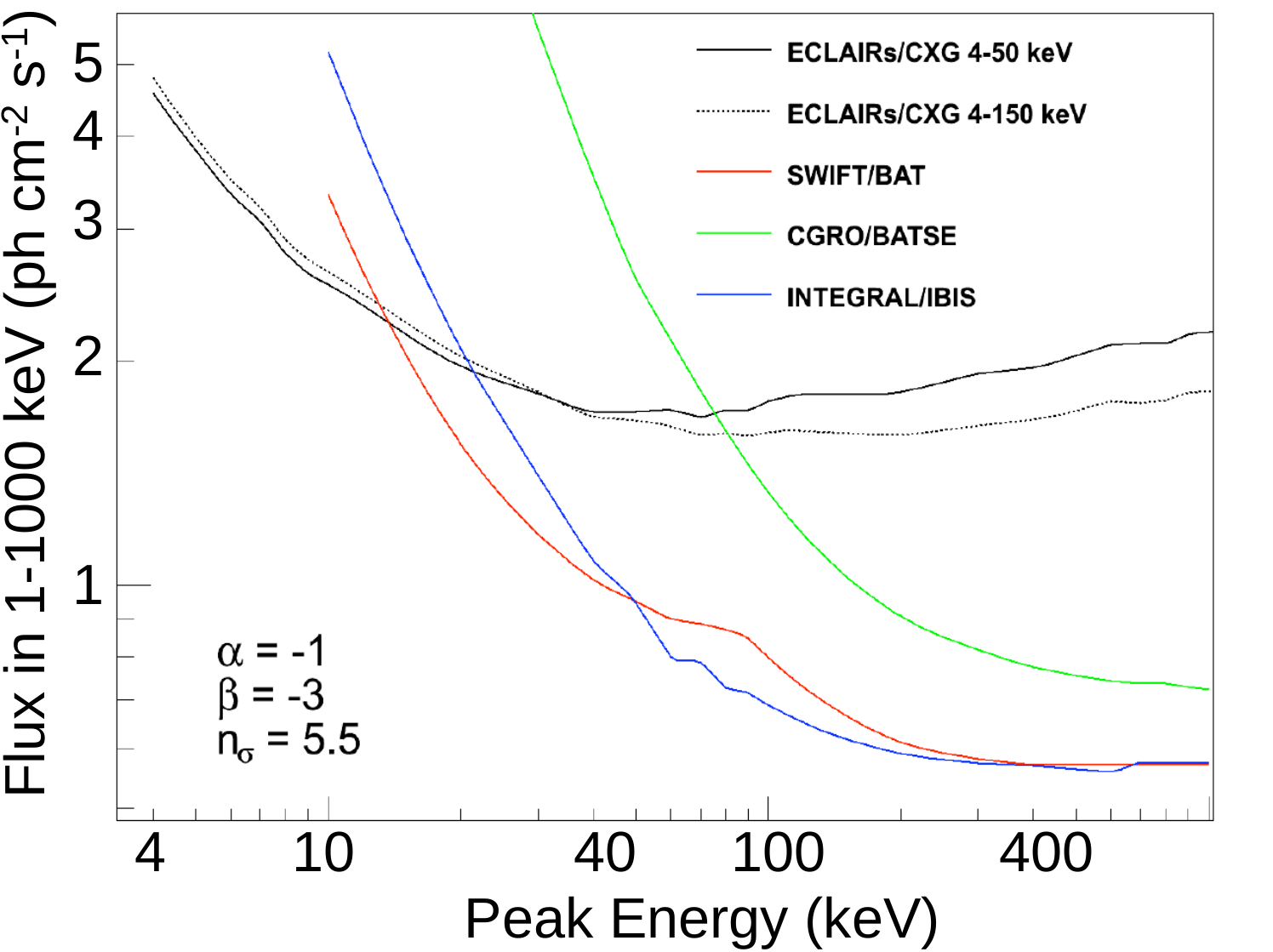}
\caption{Left: CXG mechanical sketch with the coded mask on the top, the passive lateral shielding in orange, and the detector electronics at the bottom. Right: ECLAIRs/CXG sensitivity compared to previous instruments, as a function of the GRB peak energy for a 5.5 $\sigma$ detection, assuming a Band \cite{band93} spectrum.}
\label{fig:ECLAIRs}
\end{figure}

\paragraph{ECLAIRs}
% ------------------------------
	ECLAIRs \cite{schanne08,mandrou08} is a coded-mask telescope (project lead by CEA Saclay) operating in the 4--250 keV energy range (CXG), coupled with a real-time data-processing system (UTS, see below). The CXG has a $\sim$2~sr field-of-view, and a fair localization accuracy ($\sim$10~arcmin error radius at 90\% C.L. for faint sources, a few arcmin for the brightest ones).
	Its detector plane consists of 80$\times$80 CdTe pixels.
	A new generation readout electronics (developed at CEA Saclay), a careful pixel selection and an optimized hybridization (by CESR Toulouse) allow to lower the detection threshold (from 15 keV reached by previous CdTe detectors) down to $\sim$4 keV. Thus the CXG, in spite of its rather small geometrical area (1024 cm$^{2}$), is more sensitive to GRB with soft spectra (potentially the most distant ones) than currently flying telescopes. 
	Fig. \ref{fig:ECLAIRs} sketches the CXG telescope and shows its predicted sensitivity as a function of the GRB peak energy.
	The ECLAIRs telescope is expected to localize about 70 GRB per year, taking into account the deadtime induced by the passages of the Southern Atlantic Anomaly (a region with increased particle-induced background) and the Earth in the CXG field-of-view.

\paragraph{ECLAIRs scientific trigger unit (UTS)}
% ------------------------------
The UTS (Unit\'e de Traitement Scientifique, developed at CEA Saclay) analyzes the ECLAIRs/CXG data stream in real-time to detect and localize GRB occurring within its field-of-view. It implements two simultaneous triggering algorithms \cite{schanne07} : (i) one based on the detection of excesses in the detector count-rate followed by imaging as trigger confirmation and localization, and (ii) one performing imaging on a recurrent time-base (better suited for long, slowly rising GRB).
	After GRB detection, its position is quickly sent via a VHF antenna to ground-receiver stations placed under the satellite track.
	The GRB position is also transmitted to the platform for an autonomous satellite slew (within five minutes) in order to bring the GRB in the narrow fields-of-view of the onboard MXT and VT telescopes.
	The UTS builds (from the data-streams of CXG and GRM, see below) and sends to ground (via VHF) additional GRB information (lightcurves and subimages), which provide additional real-time information on the trigger quality and the GRB spectral and temporal characteristics.

\begin{figure}[ht]
\includegraphics[page=5,height=.22\textheight]{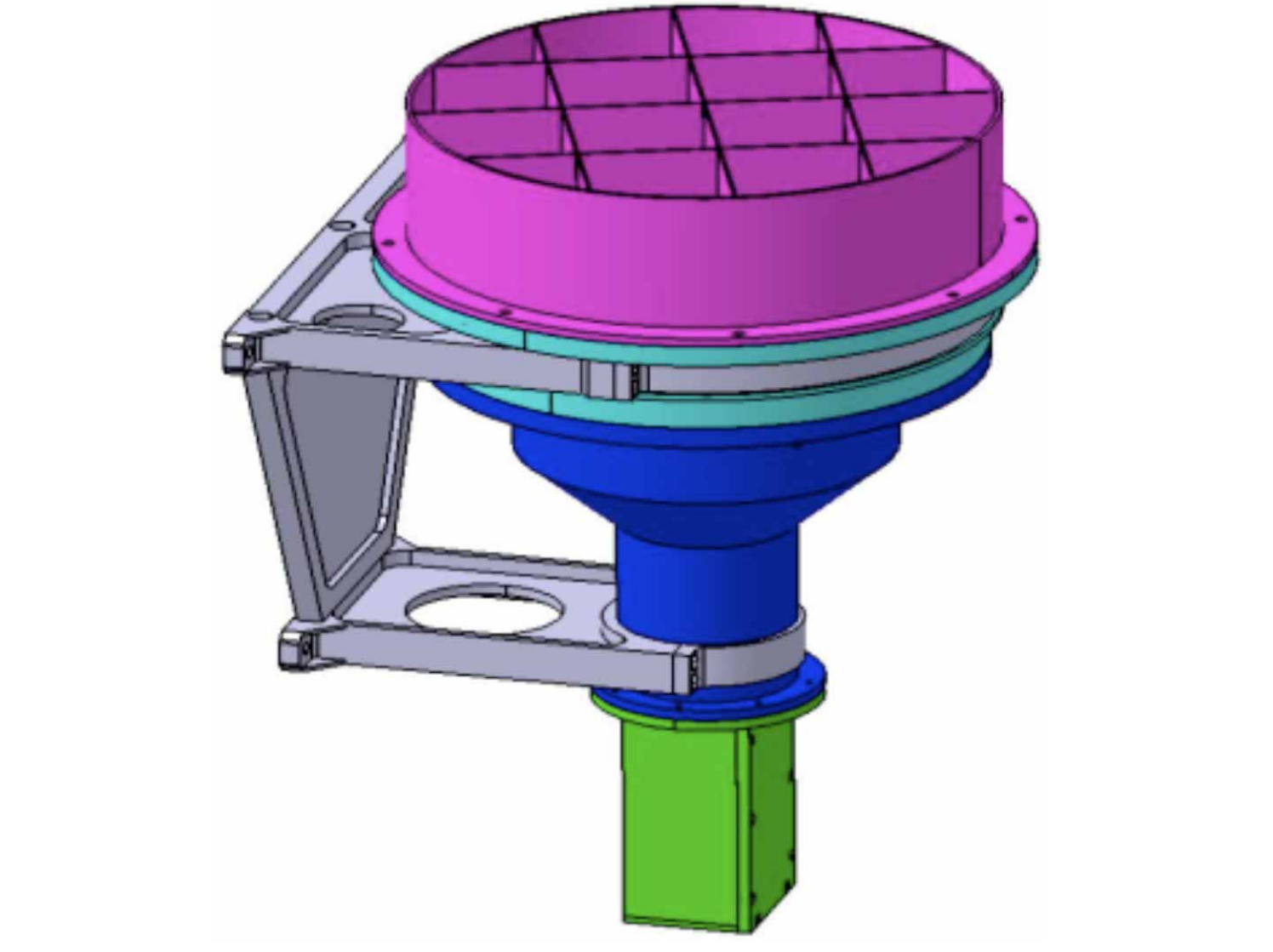}
\hspace{2cm}
\includegraphics[page=6,height=.22\textheight]{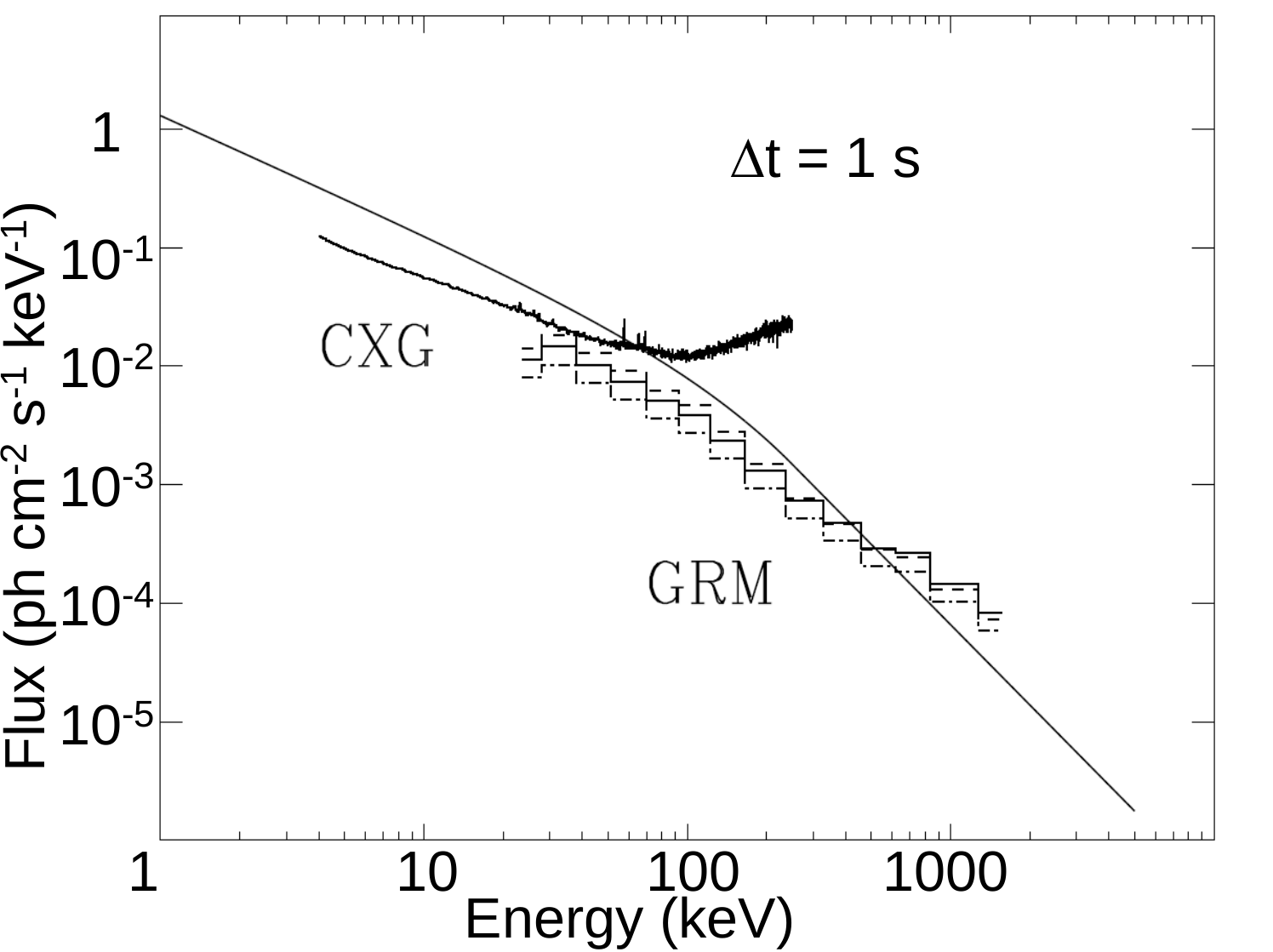}
\caption{Left: one GRM module. Right: combined GRM (two modules) and ECLAIRs on axis sensitivity (1 s integration time, 5 $\sigma$ detection). A Band spectrum ($\alpha$=--1, $\beta$=--2.5, $E_{0}$=100 keV, and F$_{50-300 keV}$=1 photon cm$^{-2}$ s$^{-1}$) is overplotted for comparison.}
\label{fig:GRM}
\end{figure}

\paragraph{GRM}
% ------------------------------
	The Gamma-Ray Monitor (GRM, developed at IHEP Beijing) on board {\it SVOM} is composed of two detector modules, each made of NaI/CsI active layers (280 cm$^{2}$ area) and a collimator for background reduction and field-of-view restriction to the one of the CXG.
	Fig. \ref{fig:GRM} shows a sketch of a GRM module and the combined CXG/GRM sensitivity.
	The GRM does not provide imaging capability, but extends the GRM spectral coverage into the MeV range, such that every {\it SVOM} GRB will have a good ECLAIRs localization and a good ECLAIRs/GRM spectrum.
	This is very important in order to determine the GRB gamma-ray spectral characteristics,
a key input for modeling the GRB's prompt radiative processes and its usage for cosmological studies.

\paragraph{MXT}
% ------------------------------
%The X-ray Imager for Afterglow Observations (XIAO) will be provided by an Italian consortium, lead
%by the INAF-IASF institute in Milan. It is a focusing X-ray telescope, based on the grazing incidence (Wolter-1) technique.  It has a short focal length of $\sim$0.8 m, and a field of view  (25 arcmin diameter) adequate to cover the whole error region provided by the CXG telescope, so that after the satellite slew the GRB position should always be inside the XIAO field of view. XIAO has an effective area of about 120 cm$^{2}$ and the mirrors are coupled to a very compact, low noise, fast read out CCD camera, sensitive in the 0.5--2 keV energy range. The sensitivity of the XIAO telescope is reported
%in Fig. \ref{fig:XIAO}, and simulations based on a sample of light-curves collected with XRT
%on board {\it Swift} show that virtually all GRB X-ray afterglows are detectable by XIAO during the first hours. This means  in practice that each GRB, for which a satellite slew is performed (not all GRB can be pointed due to different
%constraints at platform level), can be localized with a $\sim$ 5 arcsec accuracy, see Fig. \ref{fig:XIAO}. Indeed the source
%localization accuracy, is linked to the number of detected photons as $k / \sqrt{N}$ where k is a constant depending on the instrument point spread function, and N is the number of detected photons, and the early afterglows will provide
%enough photons to reach the degree of positional accuracy mentioned above.
%For more details on the XIAO instrument, see \cite{mereghetti08}

\begin{figure}[ht!]
 \includegraphics[page=7,height=.21\textheight]{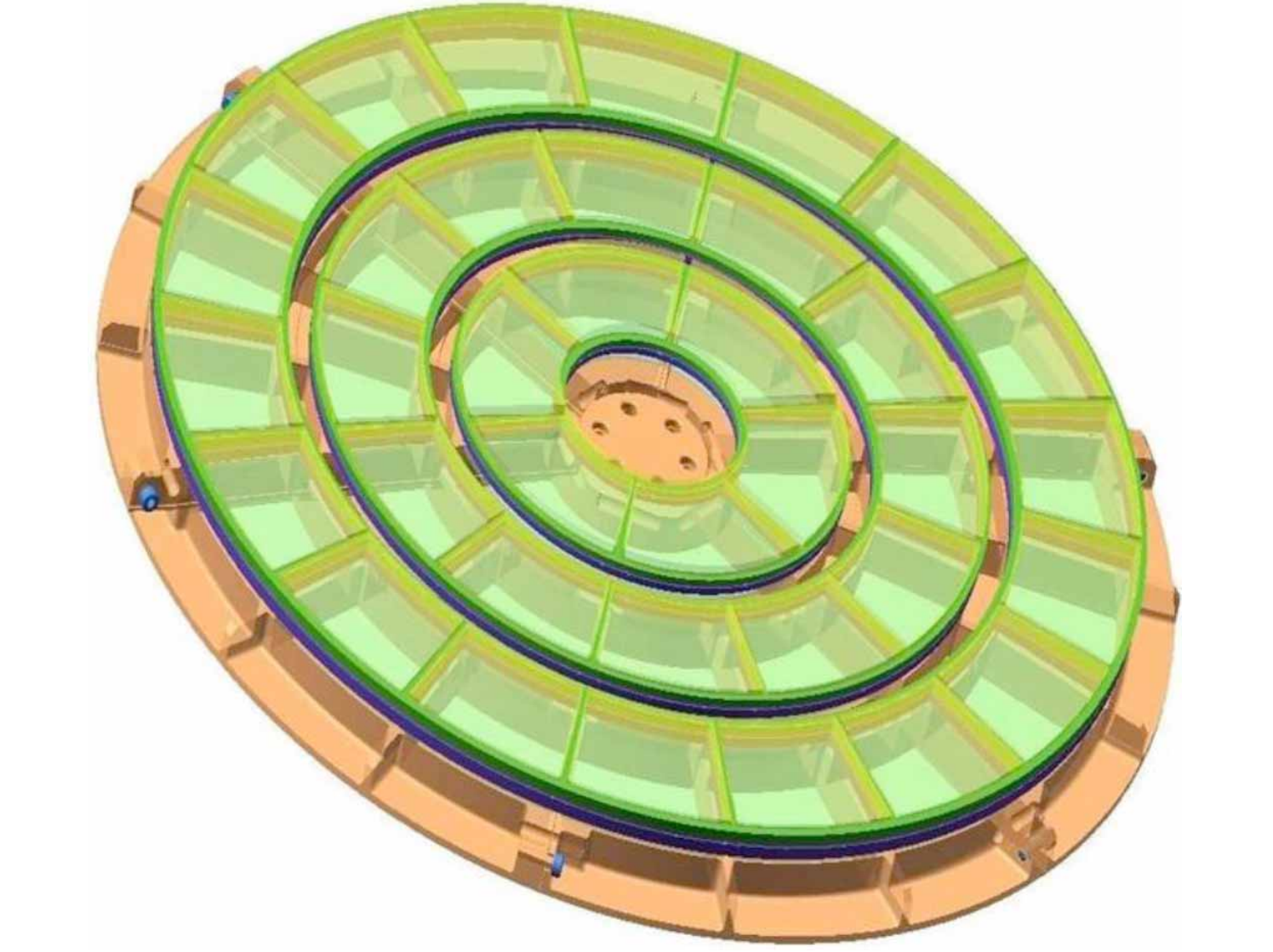}
 \hspace{2cm}
 \includegraphics[page=8,height=.21\textheight]{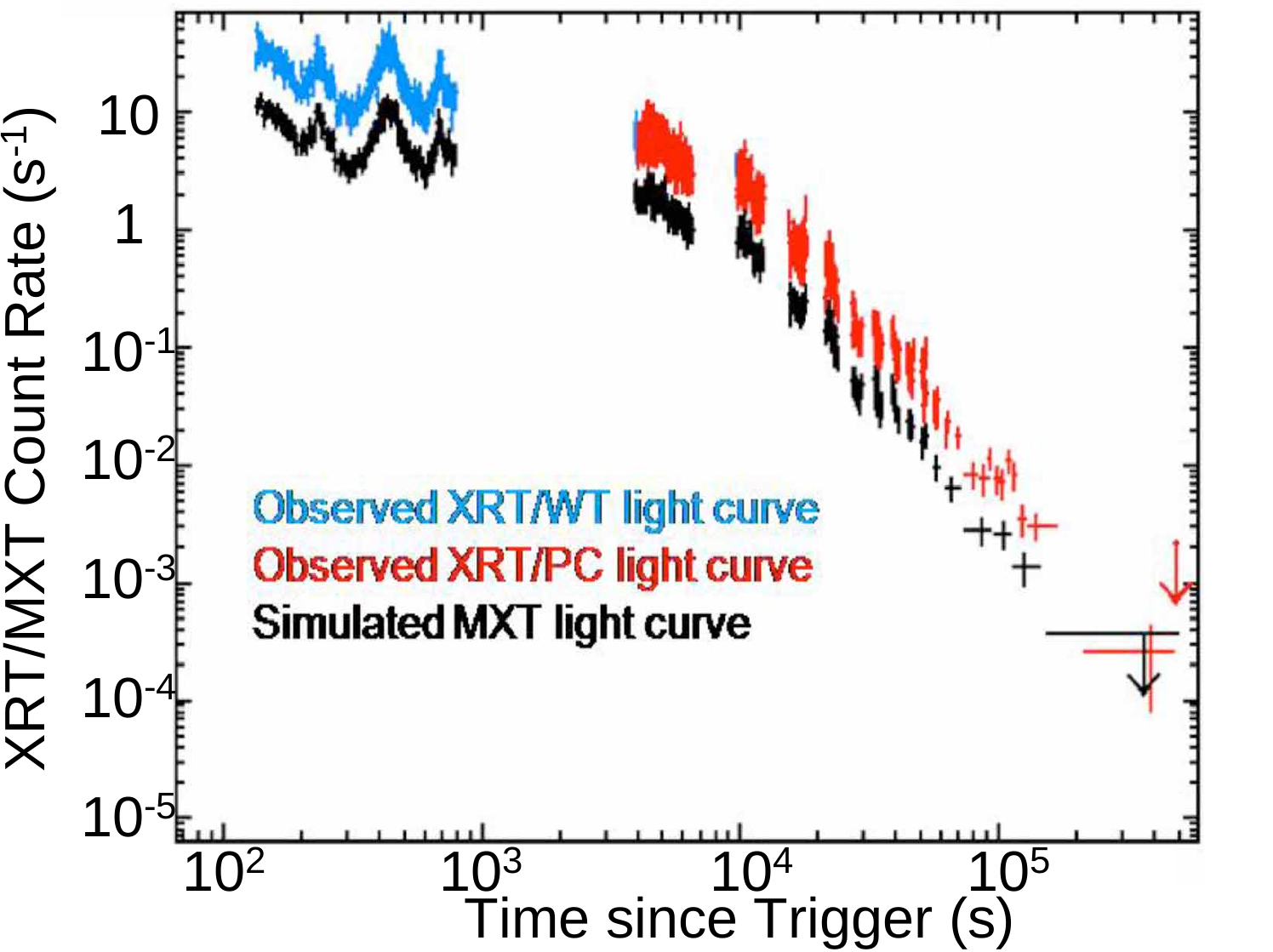}
  \caption{Left: The MXT optic module produced by assembling arrays of Microchannel plates. Right;  Predicted MXT observations compared to Swift/XRT measurements for the median brightness GRB 050730 (MXT background effect not included).}
\label{fig:mxt}
\end{figure}

	The Microchannel X-ray Telescope (MXT) for GRB X-ray afterglow observations is 
a focusing X-ray telescope, using a double microchannel-plate (MCP) reflecting optic.
	It is built by French laboratories and the University of Leicester, and is inherited from MIXS-T, an instrument to be flown on the ESA Mercury mission Bepi-Colombo \cite{mxt}.
	The MXT has a 1 m-focal length, an effective area of $\sim$50 cm$^{2}$ (at 1 keV), and a field-of-view of 1$^{\circ}$ (which covers the full ECLAIRs GRB localization-error circle) such that after any satellite slew the GRB position will be observable by the MXT.
	The MXT employs a compact, low-noise, and fast read-out CCD camera, sensitive in the 0.3--5 keV energy range.
	Fig. \ref{fig:mxt} shows a MXT simulation using a GRB light-curve acquired by Swift/XRT. A study shows that virtually all Swift/XRT GRB X-ray afterglows are detectable by the MXT during the first hours, at least half of the GRB for which a slew is performed promptly will be localized with better than $\sim$20 arcsec accuracy by the MXT within 5 minutes.

\begin{figure}[ht]
 \includegraphics[page=9,height=.21\textheight]{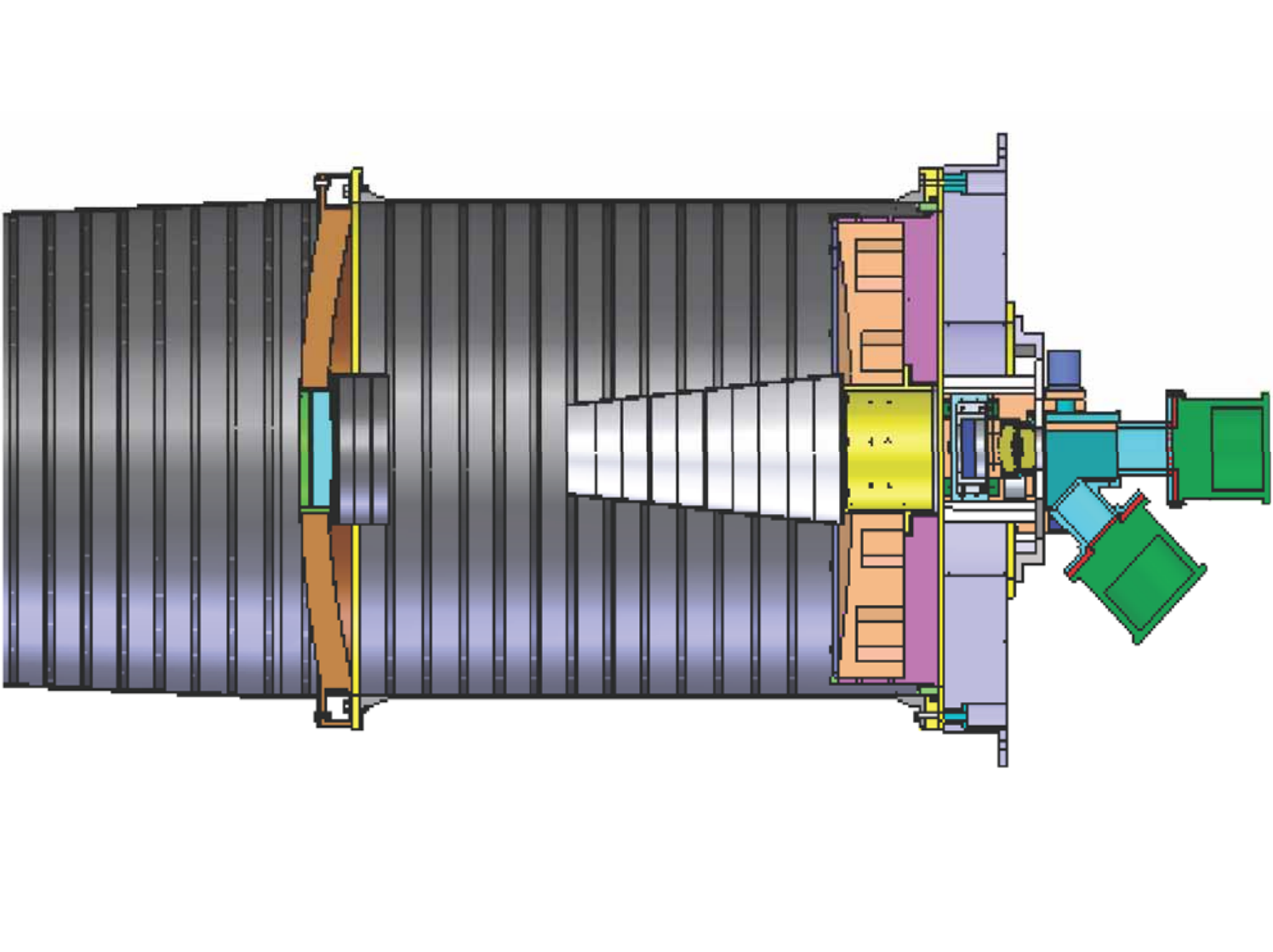}
 \hspace{2cm}
\includegraphics[page=10,height=.21\textheight]{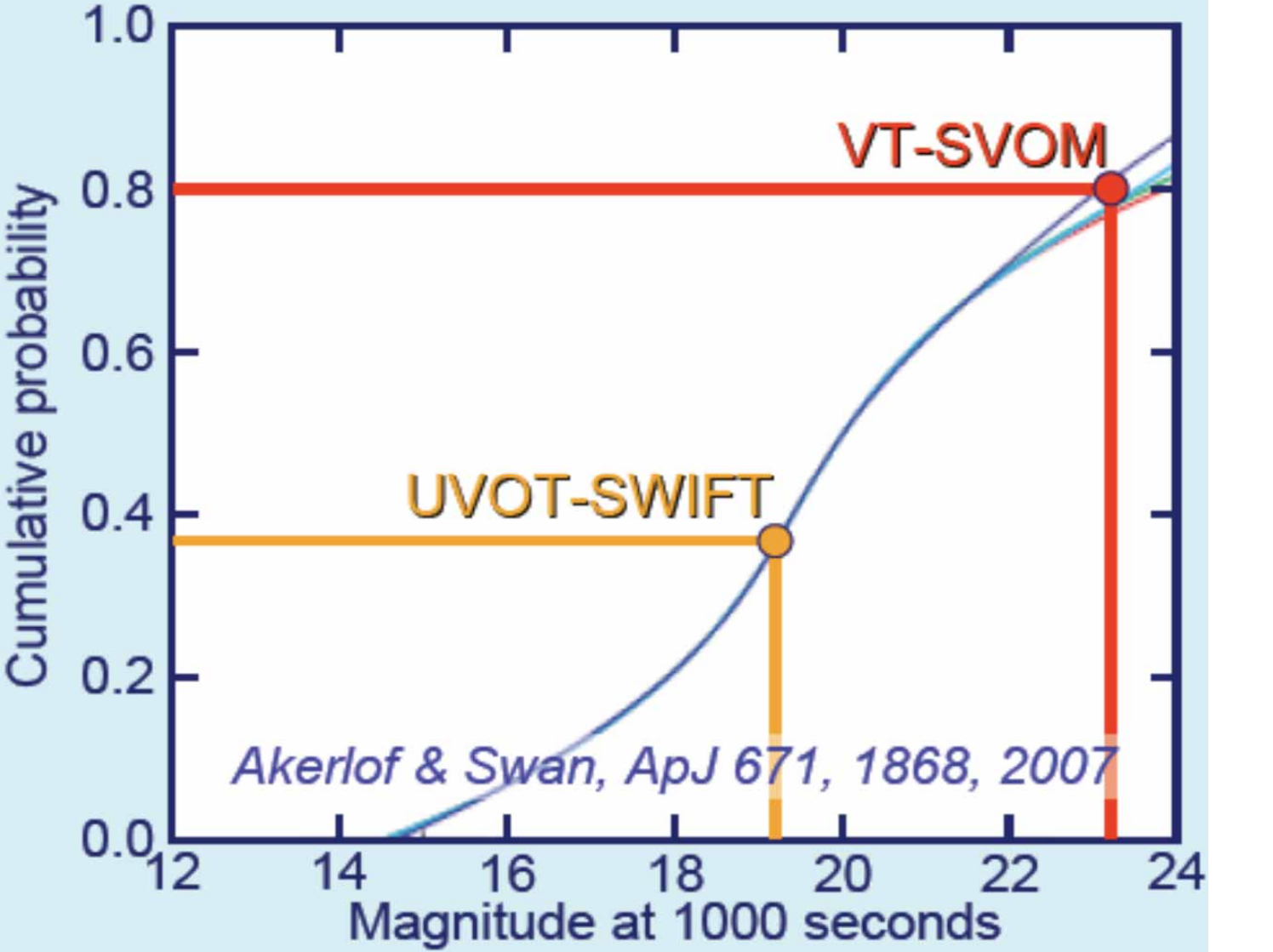}
  \caption{Right: predicted visible-band sensitivity comparison between VT (left) and UVOT (in 1000 s).}
\label{fig:vt}
\end{figure}

\paragraph{VT}
% ------------------------------
	The Visible Telescope (VT, from NAOC Beijing) improves GRB localizations to sub-arcsecond precision through optical afterglow observations, provides a uniform sample of detected optical afterglow light-curves, and allows to select optically dark GRB and high-redshift GRB candidates (z>4).
	The VT field-of-view of 21$\times$21 arcmin is adapted to the ECLAIRs/CXG error-box and its CCD images of 2048$\times$2048 pixels ensure sub-arcsecond source localizations in two spectral bands which allow to select high-redshift GRB candidates (the bands above/below 650 nm correspond to z$\sim$4-4.5, using Ly$\alpha$ absorption as redshift indicator).
	The VT provides a limiting magnitude of $M_{V}$ = 23 (5$\sigma$, 300 s exposure), a significant sensitivity improvement over current follow-up telescopes (see Fig. \ref{fig:vt}), and expects to detect $\sim$70\% of GRB for which a slew is performed.

\section{Ground segment and dedicated ground-based telescopes}
% ------------------------------
	The {\it SVOM} ground segment comprises X- and S-band antennas (for data and housekeeping telemetry
download), a mission operation center (in China), two science centers (Chinese and French, in charge of the scientific payload operations and monitoring), and the VHF alert network, which uses receiver stations distributed under the satellite track for continuous alert reception capabilities.
	The alerts contain information on refined GRB positions, sent to ground in real-time and GRB quality indicators (light-curves, images). 
	The VHF network is directly connected to the French science center \cite{claret08}, which dispatches the alerts to the scientific community through the Internet (GCN, VO Events, SVOM web page).
	The first alerts, with the ECLAIRs/UTS localization, are expected to reach the recipients one minute after the position has been computed onboard.
	The following alerts, with the X-ray afterglow position, and possibly the optical position computed from VT data, will be available after satellite slew, within 10 min.
	In case a refined (sub-arcsec) position is available from the GFTs or GWACs (see below), it is also distributed as soon as available.
	The real-time ECLAIRs pointing direction will be available in the SVOM web page, in order to allow robotic telescope pre-positioning to minimize their slew time.

%The FSC is
%in charge of the operations and monitoring of the ECLAIRs [4] payload (composed of
%the X and gamma camera CXG and the soft X-ray telescope SXT), whereas the CSC is
%in charge of the operations and monitoring of the Chinese payload (gamma-ray monitor
%GRM and visible telescope VT).

% ------------------------------
\paragraph{GWAC}
The Ground-based Wide-Angle Camera array (GWAC, from NAOC Beijing) permanently observes a portion of $\sim$8000 deg$^{2}$ of the ECLAIRs field-of-view, to search for the prompt visible emission of more than 20\% of the {\it SVOM} GRB, down to a 5$\sigma$ limiting magnitude $M_{V}$ = 15 (for a 15 s exposure) during full moon nights. About 128 camera modules are used, each with an aperture size of 15 cm, a 2048$\times$2048 pixels CCD and a field-of-view of 60 deg$^{2}$.

\paragraph{GFT} 
	Two Ground Follow-up Telescopes (GFT) automatically position their $\sim$30 arcmin field-of-view to GRB alert positions.
	In case of an optical counterpart detection, they improve the GRB localization accuracy down to 0.5 arcsec.
	Both French and Chinese GFT use multi-band optical cameras, and the French GFT has additional near infra-red capabilities.
	The telescope sites are located in tropical zones, at longitudes separated by 120$^{\circ}$ in order to reach a 40\% GRB follow-up capability.
	Low-significance alerts, considered not reliable enough to be distributed to the whole community, are also followed by those {\it SVOM}-dedicated telescopes.

\section{Conclusions}
	The future {\it SVOM} mission, equipped with instruments for GRB triggering and prompt-emission studies (ECLAIRs, GRM, GWAC) and early-afterglow follow-up telescopes (MXT, VT, GFT), will provide quick GRB characterizations (accurate positions, spectra, and photometric redshift determinations by Lyman-alpha absorption), and multi-wavelength (X-ray and visible) follow-up of 40\% of the GRB up to 10 ks after the trigger, which allows to measure the GRB spectral energy distribution during the transition between its prompt to afterglow phase.

\bibliographystyle{aipproc}   % if natbib is available

\end{document}